\pdfoutput=1 
\documentclass{JINST-Sample-files/JINST}

\usepackage{url}
\usepackage{graphicx}
\usepackage{caption}
\usepackage{subcaption}

\title{Measurements and TCAD simulation of novel ATLAS planar pixel detector structures for the HL-LHC upgrade}

\author{C. Nellist$^a$\thanks{Corresponding author.}~,
N. Dinu$^a$,
E. Gkougkousis$^a$,
and A. Lounis$^a$\\
\llap{$^a$}Laboratoire de l'Accélérateur Linéaire, CNRS\\
Bat. 200, 9140, Orsay, France.\\
E-mail: \email{clara.nellist@cern.ch}}

\abstract{The LHC accelerator complex will be upgraded between 2020--2022, to the High-Luminosity-LHC, to considerably increase statistics for the various physics analyses. To operate under these challenging new conditions, and maintain excellent performance in track reconstruction and vertex location, the ATLAS pixel detector must be substantially upgraded and a full replacement is expected.

Processing techniques for novel pixel designs are optimised through characterisation of test structures in a clean room and also through simulations with Technology Computer Aided Design (TCAD). A method to study non-perpendicular tracks through a pixel device is discussed. Comparison of TCAD simulations with Secondary Ion Mass Spectrometry (SIMS) measurements to investigate the doping profile of structures and validate the simulation process is also presented.}


\begin{document}

\section{Introduction}\label{sec:Intro}

\begin{center}
\begin{figure}[tbp]
\centering
\includegraphics[width=0.75\textwidth]{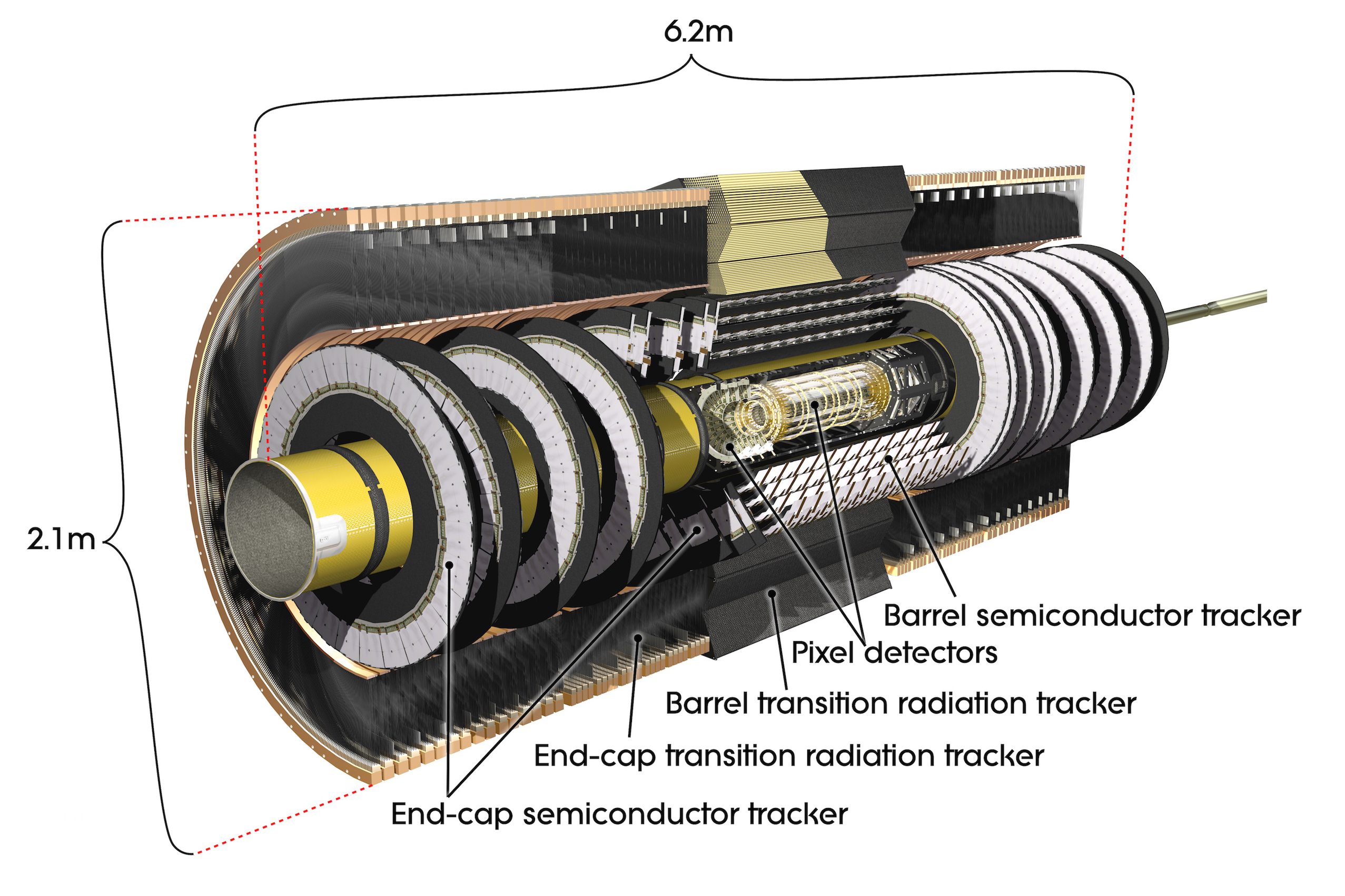}
\caption{The ATLAS Inner Detector of Run I (Image courtesy of CERN).}
\label{fig:ID}
\end{figure}
\end{center}

The Large Hadron Collider (LHC)~\cite{ref:LHC} will be upgraded to the High-Luminosity LHC (HL-LHC) during the phase II upgrade expected in $\sim$2022. These modifications will result in an increase in occupancy and of radiation damage to the ATLAS~\cite{ref:ATLAS} Inner Detector (ID), shown in figure~\ref{fig:ID}. From the interaction point outwards, the current ATLAS ID consists of the pixel sub detector, the Semi-Conductor Tracker (SCT) and the Transition Radiation Tracker (TRT). In 2014 an extra layer called the Insertable B-Layer was installed between a smaller beam pipe and the first layer of the pixel sub-detector. During the phase II upgrade, the ATLAS experiment will replace the current ID, with an all silicon tracker composing of pixels and strips. This will result in a greater area covered by pixel detectors compared to the present design. The expected fluence for the inner-most pixel layer (at approximately 4mm) will be 2~x~$10^{16}$~n$_{eq}$~cm$^{-2}$, where 1~n$_{eq}$ is 1~MeV neutron equivalent fluence. Consequently the design requires an entirely new ATLAS pixel sub-detector with improved pixel devices which are radiation hard, have slimmer edges and a better granularity.

Through new pixel designs, such as active edges and alternative bias rail geometries, minimisation of inactive regions and reductions in efficiency loss can be obtained, allowing sensors to be placed adjacent to each other. This layout is preferable to shingling as it reduces the material budget of the ID as well as cooling requirements and power consumption.

This article presents work performed in the context of this upgrade programme including the study of methods to analyse non-perpendicular particle incidences through pixel prototype devices in a clean room environment, and also the analysis of doping profiles for specifically manufactured samples with parameters used for ATLAS pixel devices.

\section{Laboratory Characterisation}\label{sec:Char}

Previous work on characterisation studies at LAL, Orsay have resulted in the redesign of the guard ring structure of planar pixel sensors, reducing the inactive region at the edge of the device from 1100~$\mu$m to 200~$\mu$m~\cite{ref:MB}.

\begin{center}
\begin{figure}[tbp]
\centering
\begin{subfigure}[c]{.45\textwidth}
\includegraphics[width=\textwidth]{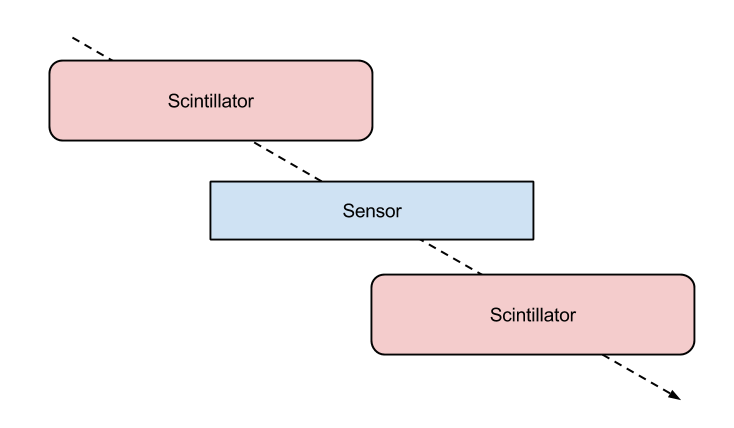}
\caption{}
\label{fig:Sketch}
\end{subfigure}
\begin{subfigure}[c]{.45\textwidth}
\includegraphics[width=\textwidth]{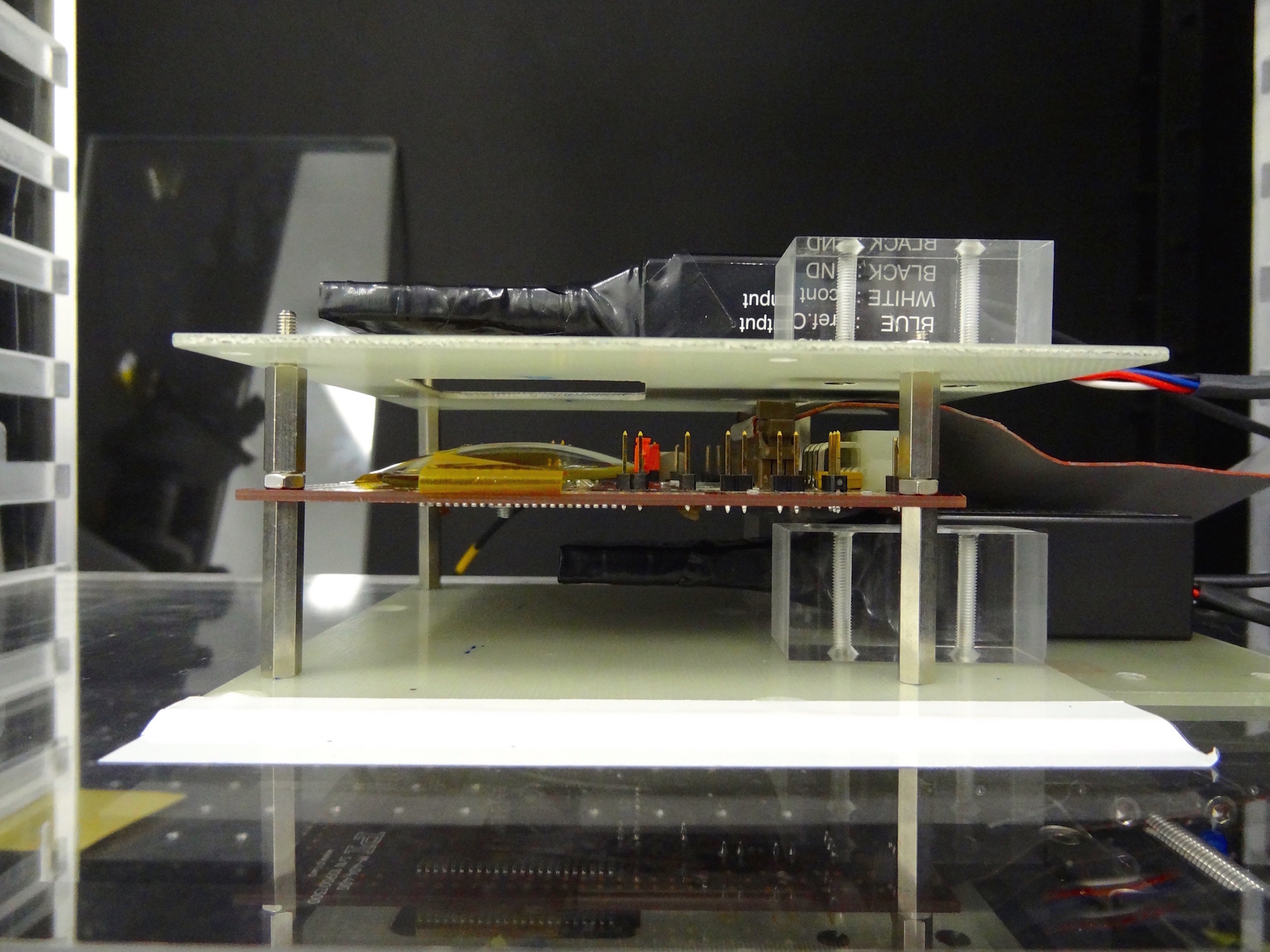}
\caption{}
\label{fig:Photo}
\end{subfigure}
\caption{(a) Sketch and (b) Photograph of the setup of offset scintillators, shown above and below the device being testes, to provide non-perpendicular tracks.}
\label{fig:Clustering}
\end{figure}
\end{center}

Devices are tested in a clean room environment with the USBPix system~\cite{ref:USBPix}, which is composed of a muti-IO board and an adapter card designed specifically for the readout-cards known as FE-I3 or FE-I4. The setup requires scintillators with photomultiplier tubes, one above and one below the device under test, for triggering on particles traversing the device either from a radioactive source, or from cosmic muons. The analogue signal from the photomultiplier tube is then fed into the wavecatcher~\cite{ref:WC} device which outputs a digital signal trigger pulse from the coincidence of the two signals. The results shown below are a proof-of-principle test on a well known device to explore the possibility of using the setup with horizontally offset scintillators, as illustrated in figures~\ref{fig:Sketch}~and~\ref{fig:Photo}, to study non-perpendicular cosmic muon tracks through the sample.

The scintillators are approximately 26~mm~by~26~mm. The first measurement taken was with a full overlap of both the scintillators with the sensor. This ``full-overlap'' case assumes that a high proportion of the tracks are perpendicular to the sensor surface, however it should be noted that angles of up to 40$^{\circ}$ from perpendicular are possible due to the size of the scintillators. Subsequently, the two scintillators were offset in the horizontal plane, in opposite directions, so that each scintillator overlapped with half of the sensor, but not with each other (this is the setup illustrated in the sketch shown in figure~\ref{fig:Sketch}). In this case, the angle of the track incidence on the sensor would be between 0$^{\circ}$ and 60$^{\circ}$. The perpendicular tracks would be rare as and only pass through the centre of the setup. Finally, the offset was increased to introduce a gap of 20~mm between the two scintillators (exactly the width of the sensor) to select tracks between approximately 34$^{\circ}$ and 67$^{\circ}$.

\begin{center}
\begin{figure}[tbp]
\centering
\includegraphics[width=0.7\textwidth]{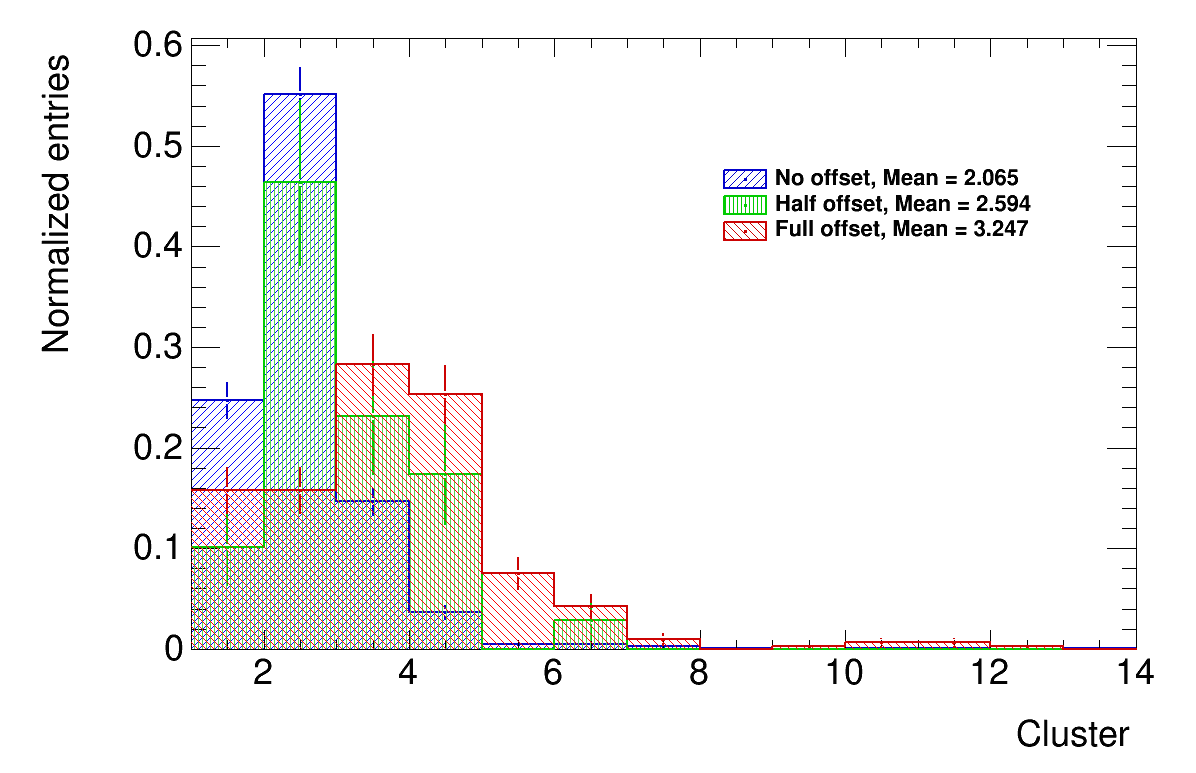}
\caption{Cluster distribution map for FE-I4 planar pixel sensor device for perpendicularly arranged scintillators (blue), half offset (green) and full offset (red).}
\label{fig:Cluster}
\end{figure}
\end{center}

The mean cluster size, as shown in figure~\ref{fig:Cluster}, increases as the scintillator gap is widened horizontally, as expected. The advantage of this method is that it can be used to study non-perpendicular tracks through the sensor when test beams are not available. When combined with cluster size information, this method could allow studies of the charge collection at various depths, and therefore various depletion regions, within the sensor bulk~\cite{ref:ChargeCol}. However, since there are fewer muons that pass non-perpendicularly and the trigger window is smaller, the method is time consuming, taking on the order of a full 24 hours to collect 1,000 events. Reconstruction of track paths is also not possible without a telescope.

\section{SIMS and TCAD Simulation}\label{sec:Sims}

\paragraph{Motivation} Results that have been obtained from test structure measurements can be used to develop reliable simulations of device, which, in turn, drive the development of new sensor layouts. This process is quicker and more cost effective than building multiple physical prototypes. It is vital, therefore, that the simulation is validated to be able to trust predictions. Studies of irradiated devices are highly important for the preparations of the ATLAS phase II upgrade and also in the context of the CERN RD50 programme~\cite{ref:RD50}, but the simulation must be first validated for non-irradiated samples.

\paragraph{}Secondary Ion Mass Spectrometry (SIMS)~\cite{ref:SIMS} is a process by which the impurities in the surface, and near surface region, of a sample can be measured~\cite{ref:Dinu}. These impurities are intentionally implanted into a silicon wafer to produce a doped region. For n-doping, impurities such as phosphorus are used, where as boron is commonly used for p-type doping. The SIMS measurement is performed by sputtering a primary energetic ion beam onto the sample and measuring the produced ionised secondary particles via mass spectrometry. This process is destructive, leaving a crater in the sample, and therefore specific samples produced for this analysis are usually required. The machine used for this study was the Cameca IMS 7F system, at the GEMAC laboratory in Versailles; further details can be found in reference~\cite{ref:Cameca}.

\begin{center}
\begin{figure}[tbp]
\centering
\includegraphics[width=0.45\textwidth]{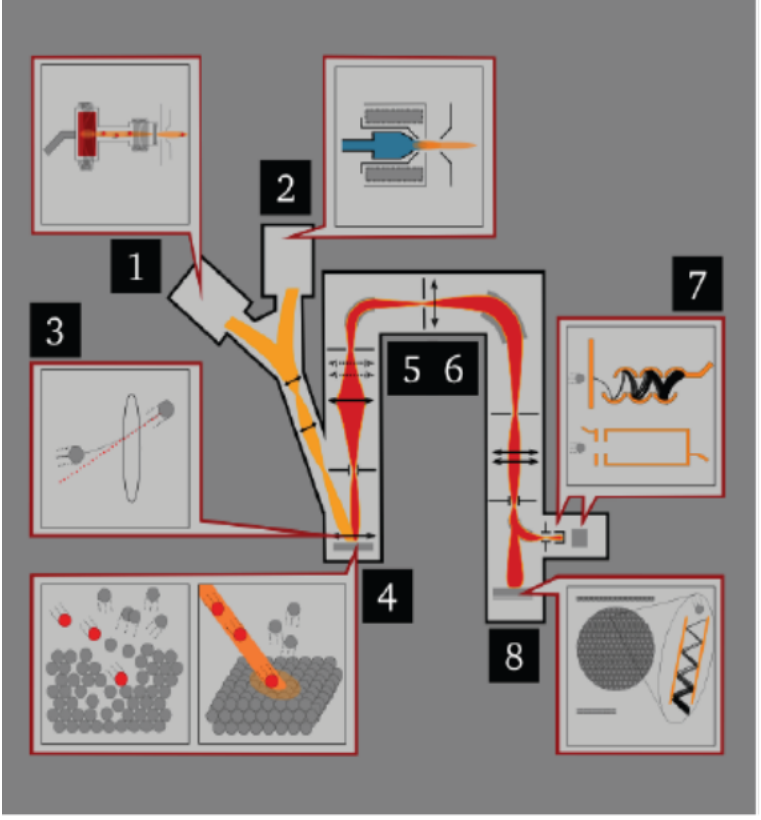}
\caption{Diagram of the SIMS process. The details of each stage, indicated by the numbers, are described in the text.}
\label{fig:SIMsDiagram}
\end{figure}
\end{center}

The SIMS process is shown in figure~\ref{fig:SIMsDiagram}. The primary ion source comes either from a duoplasmatron~(1) or from a surface ionisation source (2). A primary beam mass filter is within the primary ion column (3), which leads to the secondary ion extraction transfer (4). The ion energy is analysed at (5), the mass at (6) and the secondary ion extractors are at (7) and (8). The count rate of the elements chosen for analysis is measured as a function of time, which is subsequently converted to depth profiles with the use of a mechanical stylus. For further details of the analysis process see reference~\cite{ref:DN-HDR}.

It should be noted that this method measures the total atomic dopant profiles only, which is the physical concentration of impurities in the sample; it does not give information about the active dopant profile, that which actively contributes to the electrical field within the sample. Therefore, the total dopant profile measured by the SIMS method is not expected to change after a device has been irradiated, but it is important that this is verified.

\subsection{SIMS Sample Preparation} Samples are prepared by dicing uniformly n- or p-doped, non-thinned wafers of 6- or 8-inches with a diamond saw. The circular wafer is diced into four quadrants with four strips taken from the centre. Three quadrants are kept for the future in case further analysis should be performed. There is a layer of oxide present on the top of the sample; this layer is an insulator. The annealing process can sometimes cause an increase in the thickness of this oxide layer. Measurements are taken either through the oxide, or chemical oxide etching is performed to remove this layer. Samples for the study were manufactured at two facilities; n-in-n samples were produced at CiS, Germany~\cite{ref:CIS} and n-in-p samples were manufactured at VTT, Finland~\cite{ref:VTT}, both with a design oxide thickness of 100~nm and 200~nm. High resistivity wafers corresponding to \textgreater~4~k$\Omega$~cm$^{-1}$, and low resistivity of 0.25~$\Omega$~cm$^{-1}$, were selected. Four implantation doses ranging from $10^{13}$~at.~cm$^{-2}$ to $10^{16}$~at.~cm$^{-2}$ were chosen for study, with two implantation energies, 130~keV and 240~keV. These values were chosen because they are close to the ATLAS pixel detectors fabrication parameters. Tables~\ref{table:SIMS-CiS}~and~\ref{table:SIMS-VTT} show the two production runs for the respective manufacturers for the 100~nm and low resistivity batches only. It should be noted that VTT used a different annealing process to CiS, however the details cannot be revealed due to confidentiality agreements.

The samples with 200~nm oxide layer and high resistivity which were also produced for this analysis process, are for a further study and therefore will not be presented here.

\begin{table}[tbp]
\begin{center}
\caption{Samples manufactured at CiS for analysis with the Secondary Ion Mass Spectrometry process. Various implantation doses and energies for the samples were chosen, with the aim of studying the total doping profile in the near surface region.} 

\begin{tabular}{ l | c | c | c | c | c | c | c | c }
\hline
\multicolumn{9}{c}{n-in-n, CiS production, \textless100\textgreater~orientation, thickness 380~$\mu$m} \\ \hline \hline
Oxide thickness & \multicolumn{8}{c}{100 nm} \\
P implantation doses (atoms) & \multicolumn{2}{c |}{10$^{13}$~cm$^{-2}$} & \multicolumn{2}{c |}{10$^{14}$~cm$^{-2}$} & \multicolumn{2}{c|}{10$^{15}$~cm$^{-2}$} & \multicolumn{2}{c}{10$^{16}$~cm$^{-2}$} \\
Implantation energy (keV) & 130 & 240 & 130 & 240 & 130 & 240 &  \multicolumn{2}{c}{130}  \\
Annealing & \multicolumn{8}{c}{4~hours, 975~$^{\circ}$C} \\ \hline
\end{tabular}

\label{table:SIMS-CiS}
\end{center}
\end{table}

\begin{table}[tbp]
\begin{center}
\caption{Samples manufactured at VTT for analysis with the Secondary Ion Mass Spectrometry process. Various implantation doses and energies for the samples were chosen, with the aim of studying the total doping profile in the near surface region.} 

\begin{tabular}{ l | c | c | c | c | c | c | c | c }
\hline
\multicolumn{9}{c}{n-in-p, ADVACAM production, \textless100\textgreater~orientation, thickness $\leq$ 675~$\mu$m} \\ \hline \hline
Oxide thickness & \multicolumn{8}{c}{100 nm} \\
P implantation doses (atoms) & \multicolumn{2}{c |}{10$^{13}$~cm$^{-2}$} & \multicolumn{2}{c |}{10$^{14}$~cm$^{-2}$} & \multicolumn{2}{c |}{10$^{15}$~cm$^{-2}$} & \multicolumn{2}{c}{10$^{16}$~cm$^{-2}$} \\
Implantation energy (keV) & 130 & 240 & 130 & 240 & 130 & 240 & 130 & 240  \\
Annealing & \multicolumn{8}{c}{3~hours, 1000~$^{\circ}$C} \\ \hline
\end{tabular}
\label{table:SIMS-VTT}
\end{center}
\end{table}

\subsection{TCAD Simulation}

Simulation is performed with Technology Computer Aided Design (TCAD). Two versions have been used: Silvaco~\cite{ref:Silvo} and Synopsys~\cite{ref:Synop}. The Silvaco diffusion models considered are the following: Fermi model; 4 CPL model (clustering consideration); and PLS Solid State model. For Synopsys: Charged Pair Model; Charged Fermi model; Constant model; Charge React model. The details for each model can be found in the respective user guides. The models use varying numbers and complexities of equations to compute the diffusion process within the simulation. Models are chosen for the appropriate level of accuracy and computational cost. As such, for Synopsys, the Constant model assumes a fixed diffusivity with no electric field and is therefore the least computationally expensive, while the Charge React model solves seven equations and has a higher computational cost. Simulations with Silvaco have been performed, but will not be presented here.

The parameters for oxidation are included in the simulation~\cite{ref:Ox}, instead of adding a layer of oxide on top manually, and the simulation steps are taken from the manufacturer's own process. 

\subsection{Results}

\begin{center}
\begin{figure}[tbp]
\begin{subfigure}[c]{.5\textwidth}
\includegraphics[width=\textwidth]{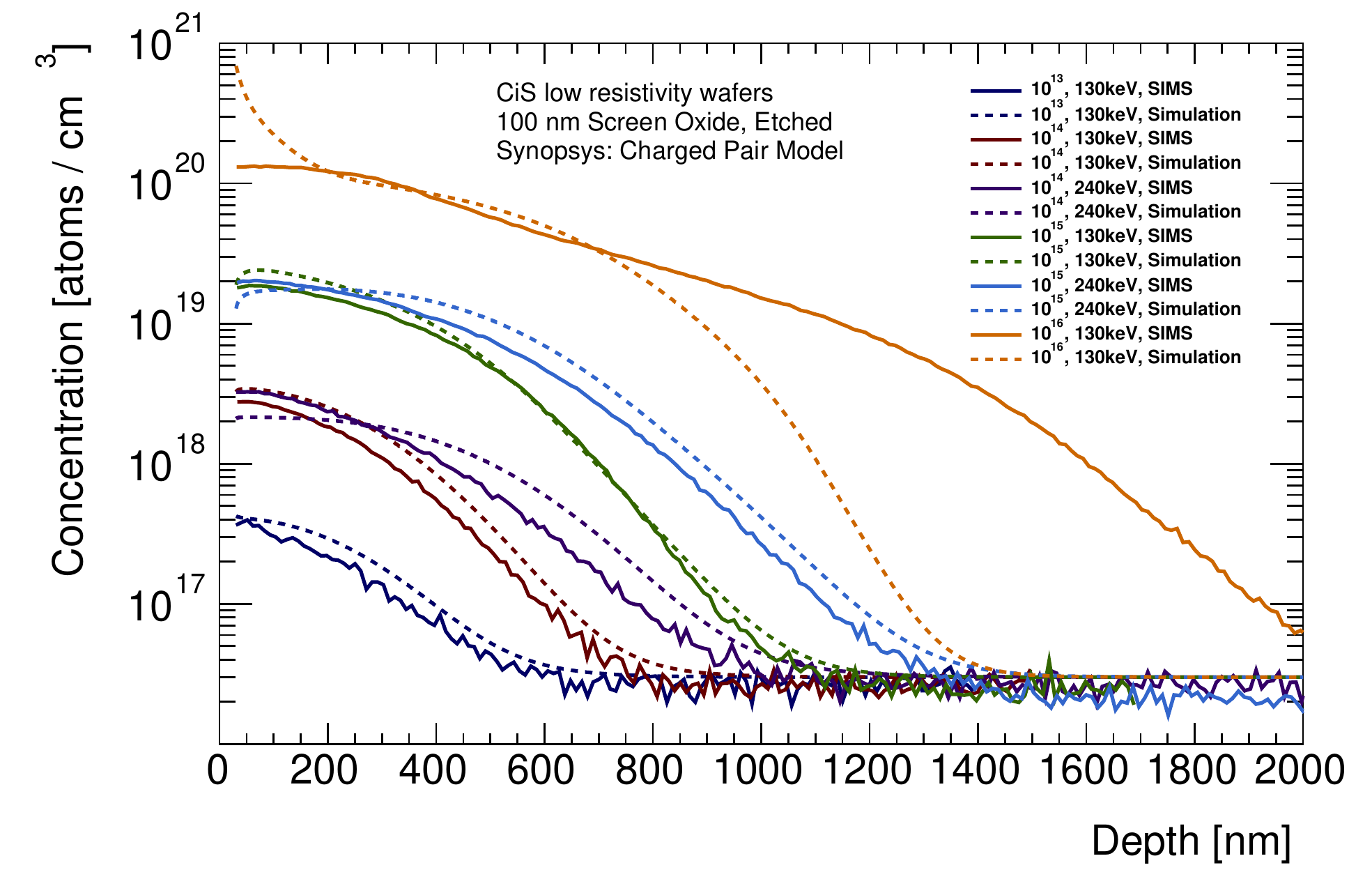}
\caption{}
\label{fig:SIMS:CP-Cis}
\end{subfigure}
\begin{subfigure}[c]{.5\textwidth}
\includegraphics[width=\textwidth]{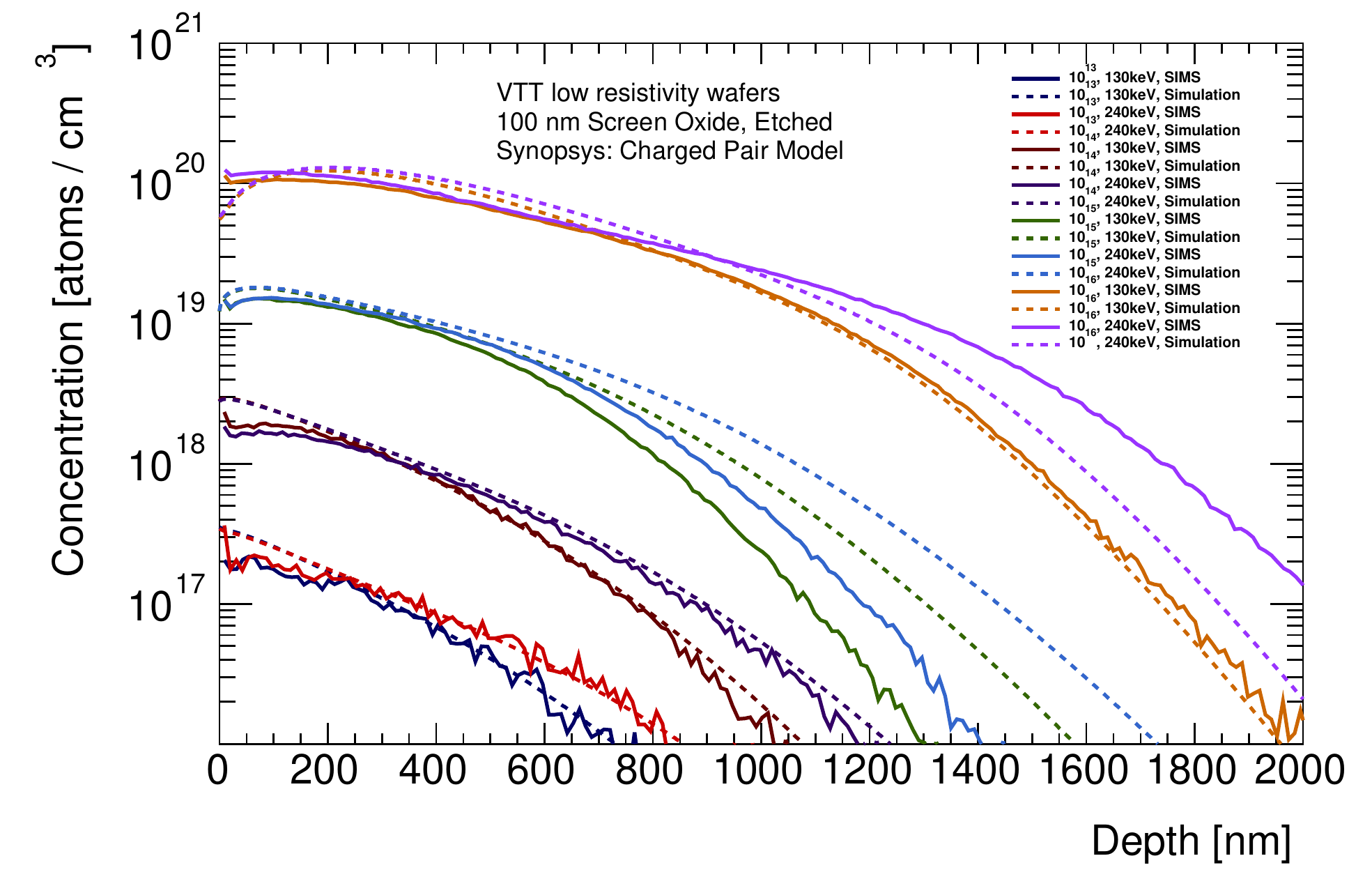}
\caption{}
\label{fig:SIMS:CP-VTT}
\end{subfigure}
\caption{TCAD simulations performed for various implantation doses (at.~cm$^{-2}$) and energies (keV) for the Synopsys Charged Pair model, compared to SIMS measurements of (a) CiS and (b) VTT low resistivity samples.}
\end{figure}
\end{center}

Simulation using the Synopsys Charged Pair diffusion model, and the SIMS data for the CiS low resistivity wafers batch have been compared, as can be seen in figure~\ref{fig:SIMS:CP-Cis}. There is generally a good agreement between them, however, a large discrepancy at the highest dose (10$^{16}$~at.~cm${-2}$) at greater depths has been observed.

\begin{center}
\begin{figure}[tbp]
\begin{subfigure}[c]{.5\textwidth}
\includegraphics[width=\textwidth]{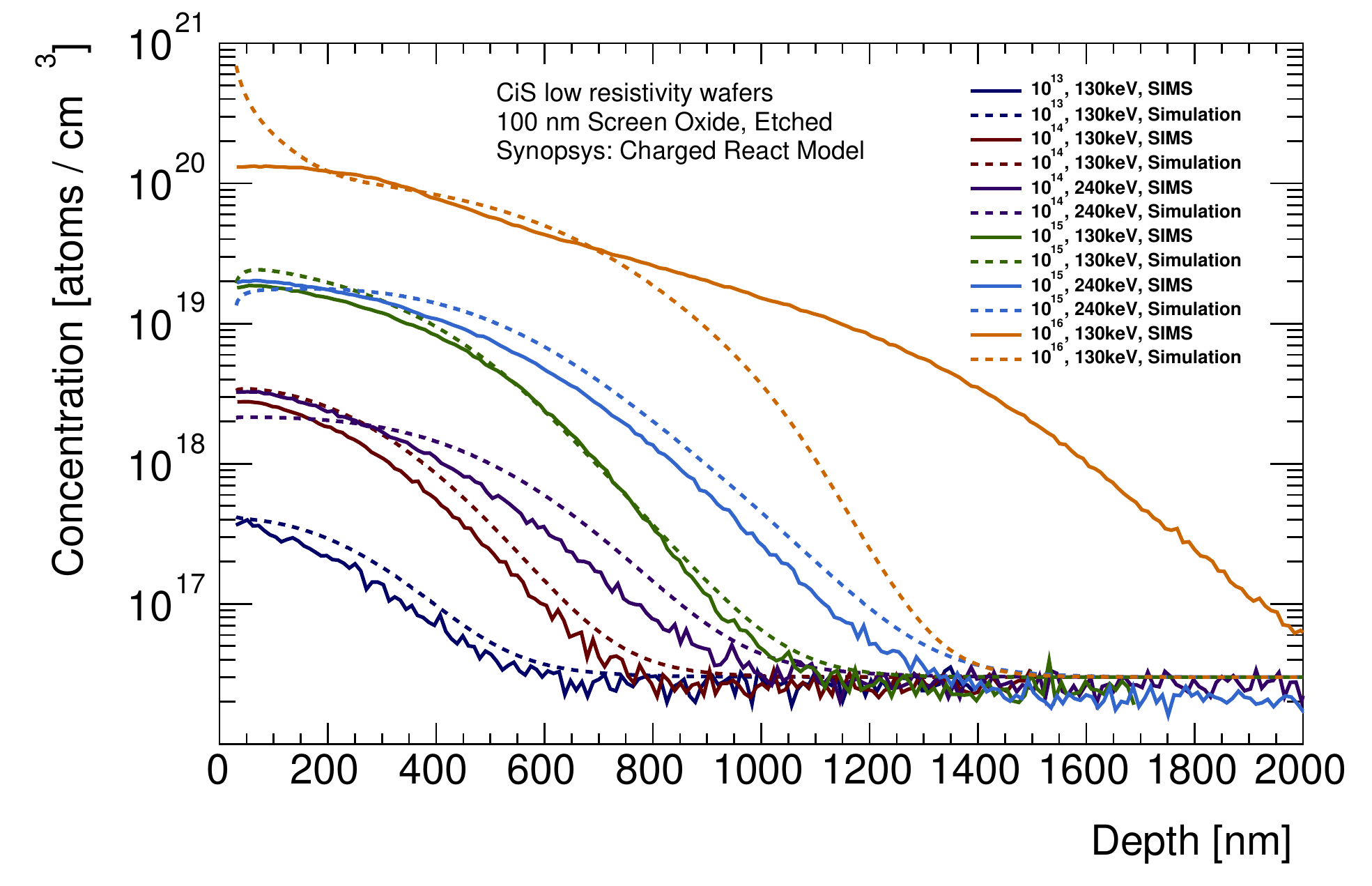}
\caption{}
\label{fig:SIMS:CR-Cis}
\end{subfigure}
\begin{subfigure}[c]{.5\textwidth}
\includegraphics[width=\textwidth]{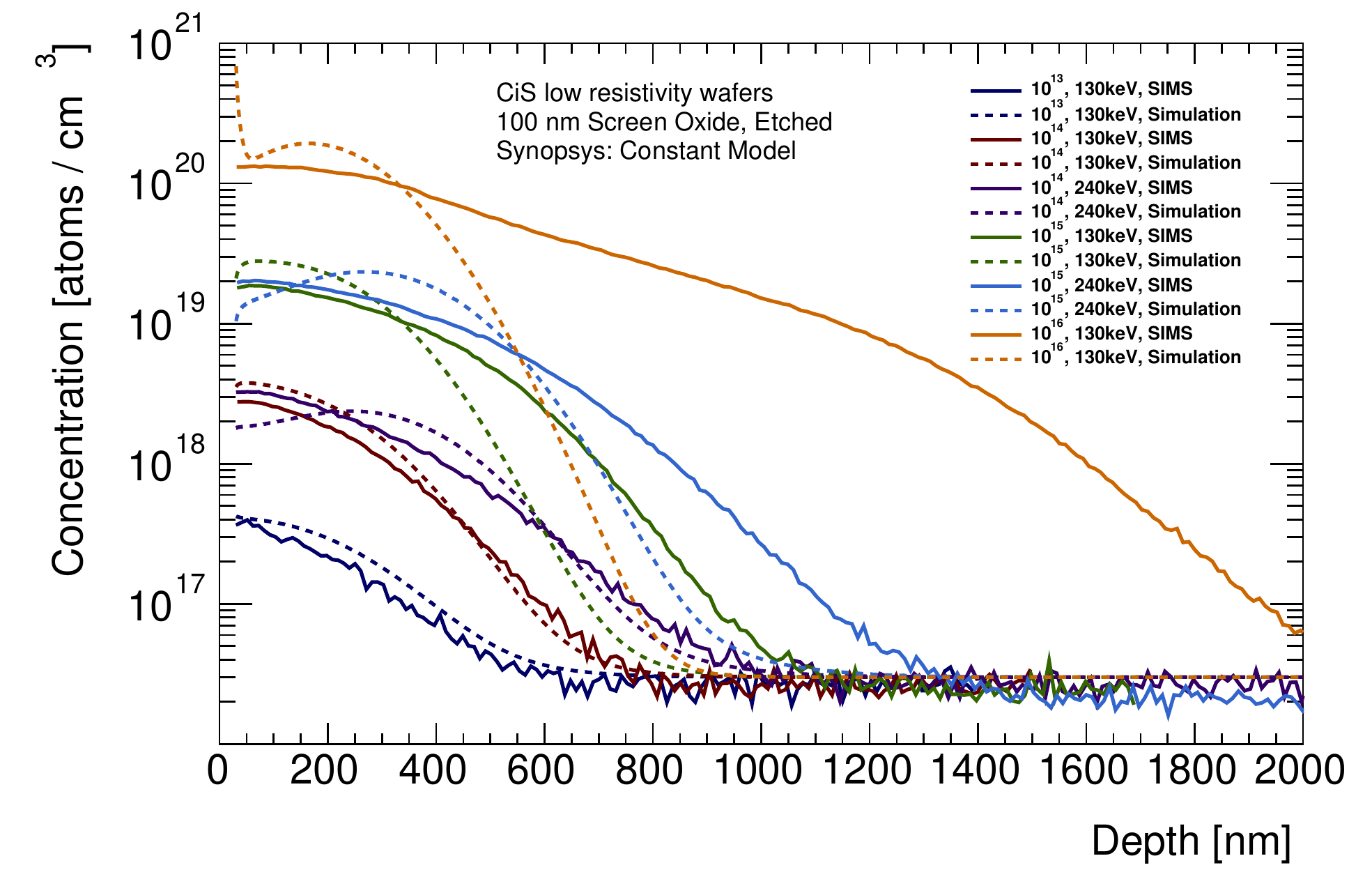}
\caption{}
\label{fig:SIMS:C-CiS}
\end{subfigure}
\caption{TCAD simulations performed for various implantation doses (at.~cm$^{-2}$) and energies (keV) for (a) Charged React and (b) Constant models, compared to SIMS measurements of CiS low resistivity samples.}
\end{figure}
\end{center}

The same comparison was performed for VTT low resistivity wafers, shown in figure~\ref{fig:SIMS:CP-VTT}. Here the discrepancy for the highest dose is not observed. To study this, various diffusion models were used, such as the Charge React and the Constant model for the CiS sample in figures~\ref{fig:SIMS:CR-Cis} and \ref{fig:SIMS:C-CiS} respectively. The discrepancy for the two CiS doping profile samples of 10$^{15}$~at.~cm$^{-2}$ is understood to be due to a systematic issue, rather than an error with the simulation, but further study is required.

\begin{center}
\begin{figure}[tbp]
\begin{subfigure}[c]{.5\textwidth}
\includegraphics[width=\textwidth]{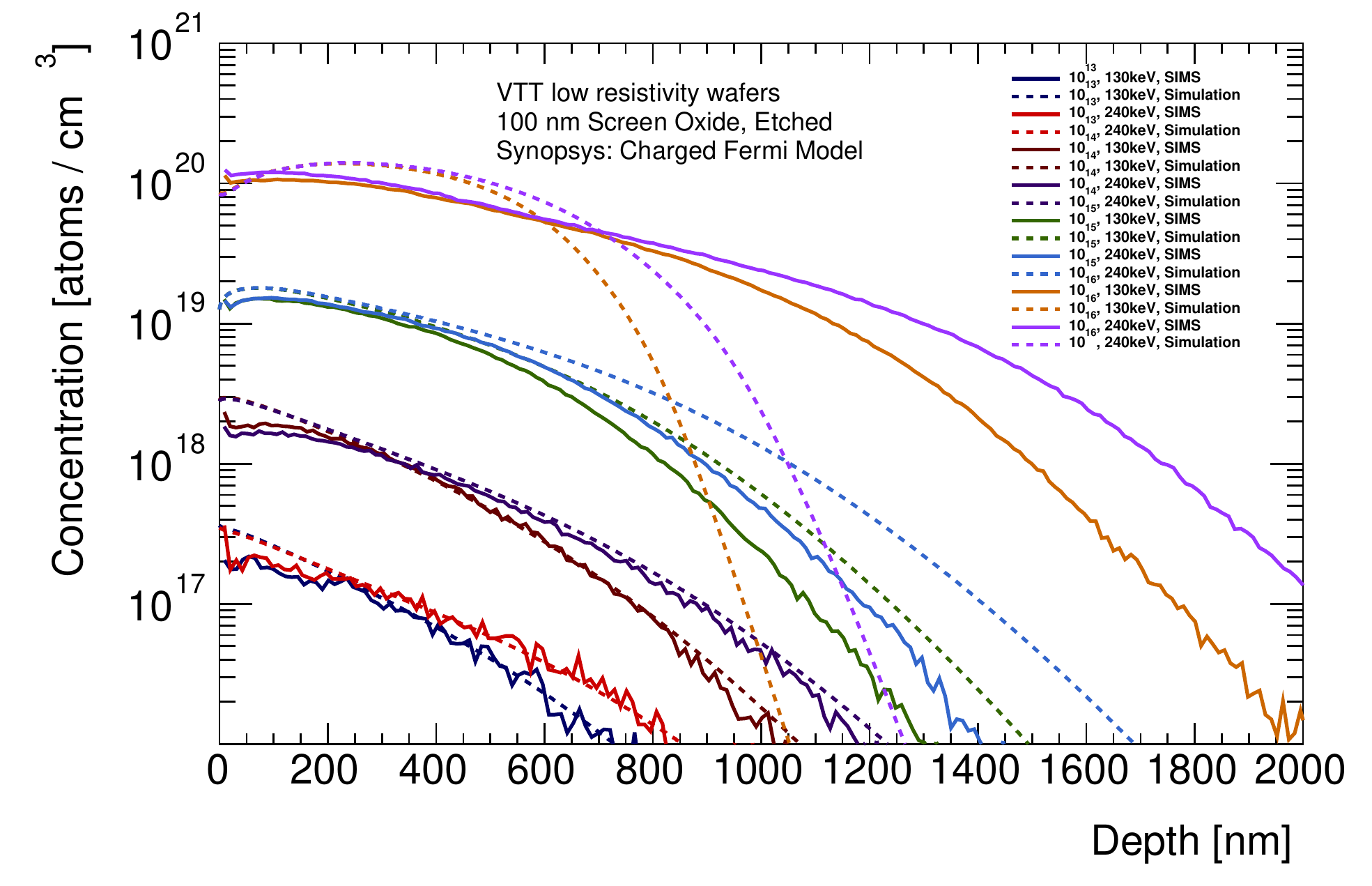}
\caption{}
\label{fig:SIMS:CF-VTT}
\end{subfigure}
\begin{subfigure}[c]{.5\textwidth}
\includegraphics[width=\textwidth]{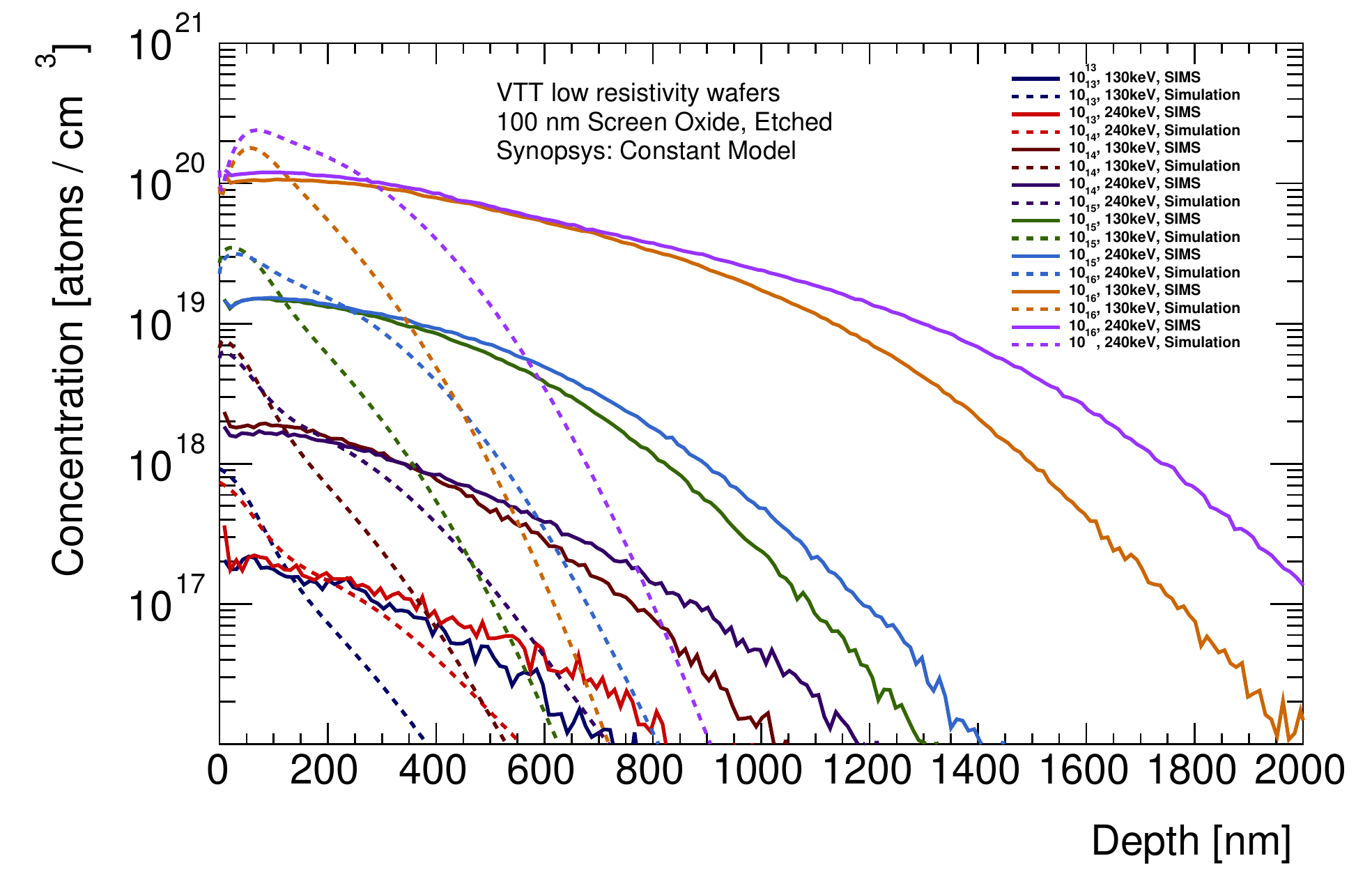}
\caption{}
\label{fig:SIMS:C-VTT}
\end{subfigure}
\caption{TCAD simulations performed for various implantation doses (at.~cm$^{-2}$) and energies (keV) for (a) Charged Fermi and (b) Constant models, compared to SIMS measurements of VTT low resistivity samples.}
\end{figure}
\end{center}

Figures~\ref{fig:SIMS:CF-VTT} and \ref{fig:SIMS:C-VTT} show the same comparison for VTT samples as described above for CiS, however results for the Charged React model are not shown due to similarities with the Charged Pair model and consequently the Charged Fermi model was studied. The comparison for the Charged Fermi model was poorer for the highest dose, and for the Constant model, all doses exhibited a significantly worse comparison. It was shown that for low and intermediate doses exhibit relatively good agreement, however for each model there is still a discrepancy for the dose of 10$^{16}$~at.~cm$^{-2}$.

As can clearly be seen, no model accurately simulates the SIMS data. Further work is required, possibly to merge two models together, combining the low depth and high depth simulations.

\vspace{-20mm}

\section{Summary and Conclusions}\label{sec:conc}

Characterisation at LAL of novel pixel prototypes for the HL-LHC, including a method to study non-perpendicular tracks through an FE-I4 sensor, has been discussed.

Results have been presented for doping profile measurements performed with the SIMS method for samples from both CiS and VTT (ADVACAM). These measurements were compared with TCAD simulations for Synopsys and Silvaco. A discrepancy has been observed between simulation and data for the highest doping profile for all simulation models for CiS structures. Further studies for 200nm oxide layer and high resistivity wafers will be
performed, and samples will also be irradiated to investigate what effect this will have on the total doping profile.

\acknowledgments

The authors would like to thank Francois Jomard and Sorin Dumitriu for their contribution to the preparation and measurements of the SIMS process.

This development has been supported by the French Labex NanoSaclay, under the project: ``Dopant profiles and imaging techniques in near field''.


\begin{thebibliography}{9}

\bibitem{ref:LHC}
L. Evans and P. Bryant, \emph{LHC Machine}, JINST 3 no. 08, (2008) S08001.

\bibitem{ref:ATLAS}
ATLAS Collaboration, \emph{The ATLAS Experiment at the CERN Large Hadron Collider}, JINST 3 no. 08, (2008) S08003.

\bibitem{ref:MB}
M. Benoit, PhD Thesis, LAL, 11-18 May 2011

\bibitem{ref:USBPix}
M. Backhaus et al., \emph{Development of a versatile and modular test system for ATLAS hybrid pixel detectors}, Nucl. Instrum. Meth. A 650 no. 1, (2011) 37 - 40.

\bibitem{ref:WC}
D. Breton, E. Delagnes, J. Maalmi, Using ultra fast analog memories for fast photodetector readout, NIMA, Volume 695, 11 Dec 2012, Pages 61-67

\bibitem{ref:ChargeCol}
Chiochia, V., et al., "Simulation of Heavily Irradiated Silicon Pixel Sensors and Comparison With Test Beam Measurements," Nuclear Science, IEEE Transactions on , vol.52, no.4, pp.1067,1075, Aug. 2005

\bibitem{ref:RD50}
Michael Moll, On behalf of the RD50 Collaboration, Development of radiation hard sensors for very high luminosity colliders?CERN-RD50 project, NIMA, Volume 511, Issues 1?2, 21 (2003), 97-105

\bibitem{ref:SIMS}
Benninghoven, A.; Rydenauer, F.G.; Werner, H.W.; \emph{Secondary Ion Mass Spectrometry: Basic Concepts, Instrumental Aspects, Applications, and Trends}, Wiley, New York, 1987.

\bibitem{ref:Dinu}
N. Dinu, et al., Dopant Profiles of Planar Pixel Sensors for the Upgrade of the ATLAS Inner Detector, IEEE-NSS Valence, 2011, NP3.M-13.

\bibitem{ref:Cameca}
AMETEK, Inc - CAMECA SAS. www.cameca.com

\bibitem{ref:DN-HDR}
N. Dinu, HDR LAL-13-192 October 2013.

\bibitem{ref:CIS}
CiS Forschungsinstitut für Mikrosensorik und Photovoltaik GmbH, Konrad-Zuse-Stra{\ss}e 14, 99099 Erfurt, Germany.

\bibitem{ref:VTT}
VTT Technical Research Centre of Finland, P.O. Box 1000, FI-02044 VTT, Finland.

\bibitem{ref:Silvo}
Silvaco Inc., http://www.silvaco.fr/products/tcad/index.html

\bibitem{ref:Synop}
Synopsys Inc., http://www.synopsys.com/Tools/TCAD/Pages/default.aspx

\bibitem{ref:Ox}
G. Franco, V. Raineri, F. Frisina, E. Rimini, Characterization of oxide layers grown on implanted silicon, NIMB, Volume 96, 1 March 1995, Pages 99-103

\end{thebibliography}
\end{document}